\documentclass{jfm}
\usepackage{graphicx}
\usepackage{epstopdf, epsfig}
\usepackage{blindtext}
\usepackage[colorlinks,citecolor = blue, linkcolor=red,hyperindex,CJKbookmarks]{hyperref}

\shorttitle{Convection-dominated dissolution for single and multiple droplets}
\shortauthor{K. L. Chong \textit{et al.}}
\title{Convection-dominated dissolution for single and multiple immersed sessile droplets}
\author{Kai Leong Chong\aff{1}\corresp{\email{k.l.chong@utwente.nl}},
      Yanshen Li\aff{1},
      Chong Shen Ng\aff{1},
      Roberto Verzicco\aff{1,2,3}
 \and Detlef Lohse\aff{1,4}\corresp{\email{d.lohse@utwente.nl}}}
\affiliation{
\aff{1}Physics of Fluids Group, Max Planck Center for Complex Fluid Dynamics, MESA+ Institute and J.M.Burgers Center for Fluid Dynamics, University of Twente, P.O. Box 217, 7500 AE Enschede, The Netherlands
\aff{2}Dipartimento di Ingegneria Industriale, University of Rome `Tor Vergata', Via del Politecnico 1, Roma 00133, Italy
\aff{3}Gran Sasso Science Institute - Viale F. Crispi, 7 67100 L'Aquila, Italy
\aff{4}Max Planck Institute for Dynamics and Self-Organisation, 37077 G\"ottingen, Germany}
\begin{document}

\maketitle

\begin{abstract}
We numerically investigate both single and multiple droplet dissolution with droplets consisting of lighter liquid dissolving in a denser host liquid. In this situation, buoyancy can lead to convection and thus plays an important role in the dissolution process. The significance of buoyancy is quantified by the Rayleigh number $Ra$ which is the buoyancy force over the viscous damping force. In this study, $Ra$ spans almost four decades from $0.1$ to $400$. We focus on how the mass flux, characterized by the Sherwood number $Sh$, and the flow morphologies depend on $Ra$. 

For single droplet dissolution, we first show the transition of the $Sh(Ra)$ scaling from a constant value to $Sh \sim Ra^{1/4}$, which confirms the experimental results by Dietrich \textit{et al.} (J. Fluid Mech., vol. 794, 2016, pp. 45--67). The two distinct regimes, namely the diffusively- and the convectively-dominated regime, exhibit different flow morphologies: when $Ra\geq 10$, a buoyant plume is clearly visible which contrasts sharply to the pure diffusion case at low $Ra$. 

For multiple droplet dissolution, the well-known shielding effect comes into play at low $Ra$ so that the dissolution rate is slower as compared to the single droplet case. However, at high $Ra$, convection becomes more and more dominant so that a collective plume \emph{enhances} the mass flux, and remarkably the multiple droplets dissolve faster than a single droplet. This has also been found in the experiments by Laghezza \textit{et al.} (Soft Matter, vol. 12, 2016, pp. 5787--5796). We explain this enhancement by the formation of a single, larger plume rather than several individual plumes. Moreover, there is an optimal $Ra$ at which the enhancement is maximized, because the single plume is narrower at larger $Ra$, which thus hinders the enhancement. Our findings demonstrate a new mechanism in collective droplet dissolution, which is the merging of the plumes, that leads to non-trivial phenomena, contrasting the shielding effect.
\end{abstract}


\section{Introduction}
Droplet dissolution dynamics is essential to many natural and industrial processes such as coating, self-cleaning, spraying on the surface, etc \citep{cazabat2010evaporation,bhushan2011natural,lohse2015surface}. It is also relevant to the extraction process used in drug delivery \citep{chou2015recent}. Droplet dissolution is in many ways similar to droplet evaporation which has been studied extensively over the past decades \citep{picknett1977evaporation,deegan1997capillary,popov2005evaporative,cazabat2010evaporation,gelderblom2011water,erbil2012evaporation, stauber2015lifetimes,shahidzadeh2006evaporating}, and it is also analogous to bubble dissolution or growth \citep{epstein1950stability,enriquez2014quasi}. The basis of all these physical processes is the same, namely the mass gain or loss of the bubble or droplet being proportional to the concentration gradient at the interface, with the concentration field outside the drop or bubble being determined by the advection-diffusion process.

Pioneering work by \cite{epstein1950stability} has put down the classical calculation for the diffusive growth or shrinkage of a gas bubble. In the theory, they consider a single spherical bubble dissolving in the bulk by pure diffusion, and the concept can be directly applied to the case of droplet dissolution \citep{duncan2006microdroplet}. As calculated by Epstein \& Plesset (EP), in the spherically symmetric case, the mass transfer rate $\dot{m}$ is given by
\begin{equation} \label{eq:EP}
\frac{dm}{dt}=-4\pi R^2D(c_s-c_\infty)\Big\{ \frac{1}{R} + \frac{1}{(\pi D t)^{\frac{1}{2}}} \Big\}.
\end{equation}
It depends on the droplet radius $R$, the mass diffusivity $D$, the saturation concentration on the surface of droplet $c_s$, the bulk concentration $c_\infty$ and the time $t$. However, in many circumstances, the droplets are sitting on the substrate instead of staying inside the bulk. To cope with that geometry, \cite{popov2005evaporative} has extended the EP theory to also be able to tackle \textit{sessile} droplets (with the quasi-static approximation):
\begin{equation} \label{eq:popov}
\frac{dm}{dt}=-\frac{\pi}{2}LD(c_s-c_\infty)f(\theta),
\end{equation}
where
\begin{equation} \label{eq:popovfactor}
f(\theta)=\frac{\textnormal{sin}\theta}{1+\textnormal{cos}\theta}+4\int_0^\infty\frac{1+\textnormal{cosh}2\theta\epsilon}{\textnormal{sinh}2\pi\epsilon}\textnormal{tanh}[(\pi-\theta)\epsilon]d\epsilon
\end{equation}
is the correction factor depending on the contact angle $\theta$ and $L$ is the footprint diameter of the droplet. 

In general, for droplet dissolution on a substrate, there are different dissolution modes that can lead to different dissolution dynamics, such as the constant contact angle, constant contact area, or stick-jump mode \citep{picknett1977evaporation,dietrich2015stick,zhang2015mixed,stauber2014lifetimes}.

However, the real situation encountered in daily life often differs a lot from the classical setup of an isolated single component drop in an infinite or semi-infinite domain. An example is multi-component dissolution \citep{chu2016dissolution,lohse2016towards}. For a multi-component drop dissolving in a host liquid, there is formation of Marangoni flow caused by the variation of surface tension over the droplet surface that in addition can influence the behaviour of emulsification \citep{tan2019microdroplet}. Similarly, the Marangoni flow also plays a crucial role in multi-component sessile droplet evaporation \citep{scriven1960marangoni,tan2016evaporation,kim2017solutal,diddens2017evaporating,edwards2018density,li2018evaporation,li2019gravitational}.

Other complicating factors that should also be taken into account are of geometrical nature. \cite{bansal2017confinement,bansal2017universal} studied the effect of confinement in the evaporation dynamics of sessile droplets in which they showed that regardless of the channel length, there are some universal features of the droplet's temporal evolution. \cite{xie2018lifetime} studied how the liquid layer surrounding the immersed droplet influences the dissolution time. They showed that dissolution slows down with the increasing thickness of the surrounding liquid layer. \cite{li2018entrapment} studied the dissolution of binary droplets with entrapment of one liquid by the other, from which they reveal a slowed down dissolution process, due to partial blockage of the more volatile liquid by the less volatile one.

Next to diffusive processes, convection can play a key role in droplet dissolution: When droplets made of a less dense liquid dissolve into a denser surrounding liquid, for large enough droplet, buoyancy can become dominant and the dissolution is no longer purely diffusive. An example is a large enough droplet composed of long-chain alcohols dissolving in water \citep{dietrich2016role}. As the density of alcohol-water mixtures is considerably less than that of water, the dissolution process can lead to solutal convection which can considerably shorten the lifetime of the droplet. The dimensionless parameter quantifying the significance of the buoyancy force over the viscous force is the Rayleigh number $Ra$. \cite{dietrich2016role} find that, for $Ra \geq 12.1$, regardless of the types of alcohol, the Sherwood number $Sh$, which is the non-dimensional mass flux, follows the same scaling relationship $Sh\sim Ra^{1/4}$.

Another crucial factor that affects the dissolution rate is the collective effect. When there are multiple droplets, one expects that the presence of the neighbouring droplets leads to shielding effects as indeed seen in \cite{laghezza2016collective,carrier2016evaporation,bao2018flow,wray2019competitive}. As a result, the lifetime for multiple droplets becomes longer than that for a single droplet. The shielding effect has also been studied in the case of collective microbubbles dissolution by \cite{michelin2018collective}. These authors have constructed the theoretical framework to account for such purely diffusive shielding effects, but for collective effects affected by convection, many questions remain open. \cite{laghezza2016collective} have experimentally studied collective droplet dissolution in the regime in which convection is relevant. They report that remarkably the neighbouring droplets can \emph{enhance} the mass flux because of enhanced buoyancy-driven convective flow in the bulk, but its detailed fluid dynamics of the process remains to be elucidated. This is not possible in \cite{laghezza2016collective} because the lattice-Boltzmann simulations employed in that paper do not include convection and the underlying mechanism could thus not yet be elucidated.

In this study, we investigate both single and multiple droplet dissolution by numerical simulations with convection being considered in all cases. The structure of the paper is as follows: In Section \ref{sec:method} we introduce the numerical method for simulating droplet dissolution. In Section \ref{sec:verif} we provide the code verification. We then present the results and discussions, first for a single droplet case (Section \ref{sec:resulta}) and then for multiple droplets (Section \ref{sec:resultb}). In Section \ref{sec:conc} the conclusions and a outlook are given.

\section{Numerical method and parameters}\label{sec:method}
The simulation of droplet dissolution consists of two parts. The first is the coupled solution of the velocity field $\tilde{\textbf{u}}(\bf{x},t)$, the (kinematic) pressure field $\tilde{p}(\bf{x},t)$ and the concentration field $\tilde{c}(\bf{x},t)$, using the three-dimensional Navior-Stokes equations, advection-diffusion equation and incompressible condition within the Oberbeck-Bounssinesq approximation,
\begin{equation} \label{eq:mom}
\partial_t \tilde{\textbf{u}}+(\tilde{\textbf{u}}\cdot \nabla)\tilde{\textbf{u}}=-\nabla \tilde{p}+\sqrt{\frac{Sc}{Ra}}\nabla^2\tilde{\textbf{u}}+\tilde{c},
\end{equation}
\begin{equation} \label{eq:diff}
\partial_t \tilde{c}+(\tilde{\textbf{u}}\cdot \nabla)\tilde{c}=\sqrt{\frac{1}{RaSc}}\nabla^2\tilde{c},
\end{equation}
\begin{equation} \label{eq:incom}
\nabla \cdot \tilde{\textbf{u}}=0.
\end{equation}
The two dimensionless control parameters are the Rayleigh number 
\begin{equation} \label{eq:Ra}
Ra=\frac{g\beta_c c_s R_0^3}{\nu D}
\end{equation}
and the Schmidt number 
\begin{equation} \label{eq:Ra}
Sc=\frac{\nu}{D}.
\end{equation}
Here, $g$, $\beta_c$, $c_s$, $R_0$, $\nu$ and $D$ are the gravitational acceleration, the solutal expansion coefficient, the saturation concentration of the solute, the initial droplet radius, the kinematic viscosity, and the diffusion coefficient, respectively. Equations (\ref{eq:mom}) and (\ref{eq:diff}) were already made dimensionless using the initial droplet radius $R_0$, the free fall velocity $u_{\mathrm{ff}}=\sqrt{\beta_c g c_s R_0}$ (and the corresponding time $t_{\mathrm{ff}}=R_0/u_{\mathrm{ff}}$), and the saturation concentration $c_s$, such that the dimensionless radius, radial distance, velocity, time and concentration related to the dimensional ones in the way $\tilde{R}=R/R_{0}$, $\tilde{r}=r/R_{0}$, $\tilde{\textbf{u}}=\textbf{u}/u_{\mathrm{ff}}$, $\tilde{t}=t/t_{\mathrm{ff}}$ and $\tilde{c}=c/c_s$.

\begin{figure}
  \centerline{\includegraphics[width=1.0\textwidth]{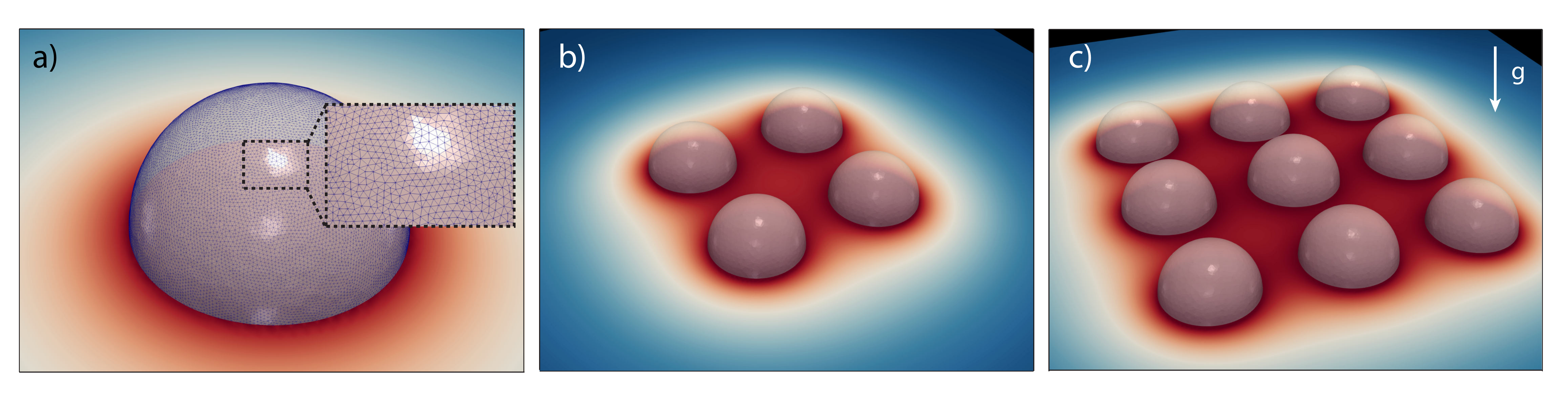}}
  \caption{(a) Schematics for triangulated Lagrangian meshes for the immersed boundary method. The configuration of multiple droplets with 2$\times$2 are shown in (b) and 3$\times$3 arrays in (c).}
\label{fig:1}
\end{figure}

The second part of the solver involves the equation that governs the dynamics of the droplet dissolution, i.e. the rate of change of the droplet radius. In this study we assume that the dissolution is in the constant contact angle mode at $90^\circ$. Therefore, the temporal change of the dimensionless droplet radius does not contain the explicit contact angle dependence and can be written as
\begin{equation} \label{eq:drop}
\frac{d\tilde{R}}{d\tilde{t}}=\frac{c_s}{\rho_d}\frac{1}{\sqrt{RaSc}} \Big\langle \frac{\partial \tilde{c}}{\partial \tilde{r}} \Big|_{\tilde{r}=\tilde{R}}\Big\rangle_S.
\end{equation}
Here $\rho_d$ and $\langle . \rangle_S$ represents the density of the droplet and the averaging over entire surface of the droplet, and $\frac{\partial \tilde{c}}{\partial \tilde{r}} \Big|_{\tilde{r}=\tilde{R}}$ is the outer concentration gradient at the boundary of the droplet.

We solve the equations using the second order finite difference method with a fractional-step third order Runge-Kutta (RK3) scheme \citep{verzicco1996finite,van2015pencil}. To impose the interfacial concentration of the immersed droplet(s), the Moving Least Squares (MLS) based Immersed Boundary Method (IBM) has been used. For this method, the boundary of each droplet is represented by a network of triangular elements (see inset of figure \ref{fig:1}a) and the movement of those elements are governed by the equation (\ref{eq:drop}), in which the concentration gradient on the surface of the droplet $\frac{\partial \tilde{c}}{\partial \tilde{r}} \Big|_{\tilde{r}=\tilde{R}}$ can be computed through interpolating the concentration at the probe locating at short distance outside the droplet. For details of our MLS-based IBM method, we refers to \cite{spandan2017parallel}.

The boundary condition at the surface of the droplet(s) is set to be the saturation concentration $c_s$ for the concentration field while it is assumed to be no-slip and no penetration conditions for the velocity field, disregarding any possible flow in the droplet. For the Cartesian container, the boundary condition for the concentration field is taken as no mass flux at all walls except the outflow boundary condition taken for the top walls. The boundary conditions for the velocity field are taken as (i) no-slip at the bottom wall (ii) periodic at the sidewalls (iii) outflow boundary condition for the top wall which is done by setting the vertical gradient of all the velocity components to be zero. It is worth to note that the advantage of using outflow boundary condition at the top walls is to minimize the finite domain size effect. It is especially useful in the situation of large $Ra$ where upward moving plumes is observed, as this outflow boundary condition prevents an artificial accumulation of solute over the domain. 

In this study, we focus on the cases of large Schmidt number, namely $Sc=1200$ as for long-chain alcohol dissolving in water, as done in the experiments of \cite{dietrich2016role}. These simulations are challenging because the mass diffusivity is much smaller than the viscous diffusivity, and thus the resolution for the scalar field is more demanding than that for the velocity field, implying that---if the same grid is used for all fields---the resolution for the most time-consuming momentum solver and pressure solver become redundant.  To overcome this challenge, we use the multiple-resolution strategy to solve the momentum and the scalar equations \citep{ostilla2015multiple}. In all cases the mesh of $144\times144\times144$ is used to resolve the velocity field, whereas the mesh for the concentration field has been doubled which is $288\times288\times288$. This mesh might appear still small for $Sc=1200$. In our case, however, we have a very small $Ra$; therefore the total P\'eclet number $Pe=\sqrt{RaSc}$, that rules the scalar diffusivity, remains smaller than $700$.

We will present the result of droplet dissolution for Rayleigh numbers spanning almost four decades ($0.1 \leq Ra \leq 400$) and for $Sc$ fixed at $1200$. As seen from equation (\ref{eq:drop}), the dynamics of the dissolution is also governed by the ratio of the saturation concentration to the density of the droplet, $c_s/\rho_d$. Here we consider the particular case where $c_s/\rho_d=0.027$, corresponding to 1-pentanol in water. 

Apart from the single droplet dissolution, we also investigate convection in the situation of multiple droplets. Two different multiple droplet configurations, namely $2\times2$ and $3\times3$ droplet arrays, have been explored. In both cases the droplet separation (measured from the edge of the droplet) equals half of the droplet initial radius (see figure \ref{fig:1} for the illustration of the set-up).

\begin{figure}
  \centerline{\includegraphics[width=1.0\textwidth]{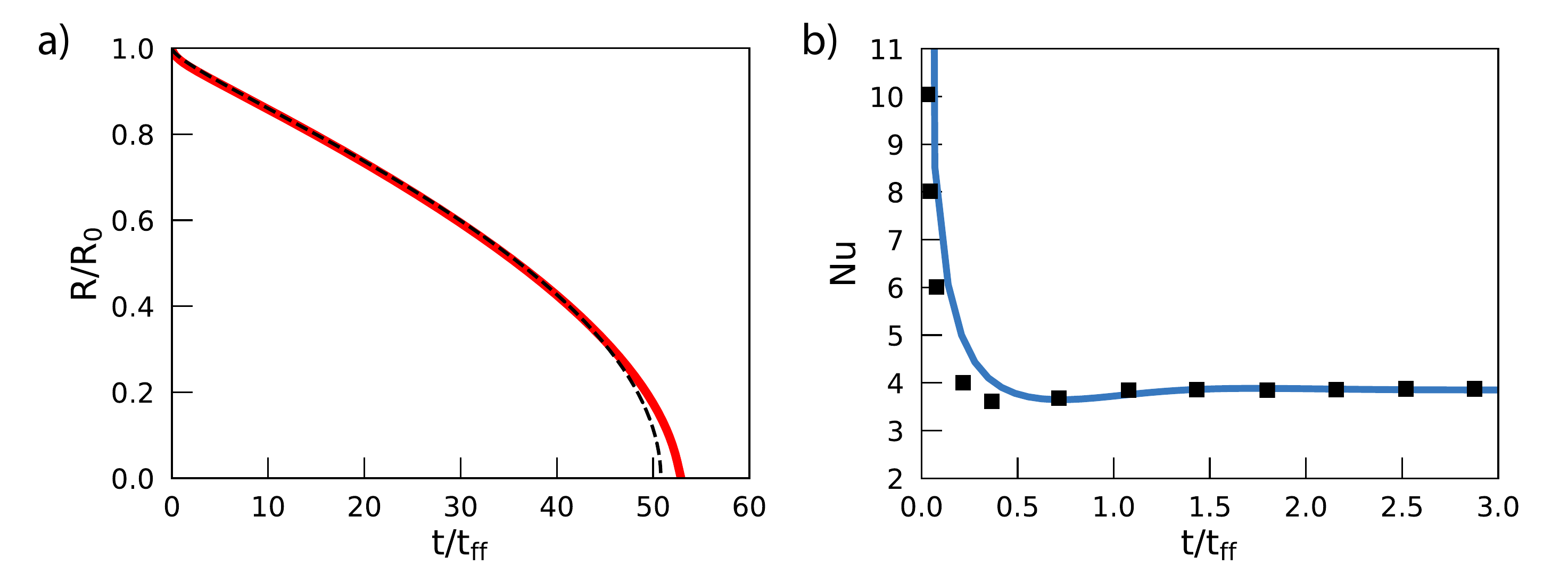}}
  \caption{(a) Numerical results (red curve) for the droplet radius as function of time for pure diffusion, compared to the Epstein-Plesset theory (black dashed curve) with the correction term proposed by \cite{popov2005evaporative}. (b) Nusselt number $Nu$ versus time $t$ for the case of a constant temperature spherical object where the black squares denote the data set given by \cite{musong2016} and the blue curve is the result from our simulation.}
\label{fig:2}
\end{figure}

\section{Code verifications}\label{sec:verif}
Before presenting the results, we first verify our code against the analytical solution and the existing results from the literature. In the first part of the verification, we consider the dissolution of a sessile droplet with pure diffusion, i.e. we solve equation (\ref{eq:diff}) with the advection term being switched off. 

\cite{epstein1950stability} considered a particular case for a single spherical bubble dissolving in the bulk fluid and analytically calculated the radius as function of time. Later \cite{popov2005evaporative} extended this calculation to the case of a droplet sitting on a substrate at a given contact angle $\theta$. However, Popov's original model assumes the quasi-static behaviour, i.e., the time dependent term in the right hand side of equation (\ref{eq:EP}) is eliminated. This assumption can largely affect the numerical dissolution process, as shown by \cite{zhu2018diffusive}. Therefore in the verification, instead of directly using the mathematical expression in equation (\ref{eq:popov}), we adopt the contact angle correction factor $f(\theta)$ as proposed by \cite{popov2005evaporative} to the classical EP theory. In the case of $\theta=90^\circ$ and a single drop considered here, this just leads to the solution given in the equation (\ref{eq:EP}) and hereafter we still call this as EP theory for simplicity. Note that due to the axial symmetry assumption in the calculation, it is only suitable for verifying the cases of a single droplet without convection but not for the cases of multiple droplets.

Figure \ref{fig:2} (a) shows the normalized droplet radius $R/R_o$ versus the dimensionless time $\tilde{t}$ at $Ra=0.01$. The figure shows the excellent agreement between the EP theory (black dashed curve) and our numerical results (red curve) over the entire dissolution process. We remark that there is a little deviation at the final stage of the dissolution because the droplet size is getting smaller and the resolution of the Eulerian grid points in the Cartesian container becomes insufficient to resolve the droplet. However, if we focus on the lifetime of the droplet, which is the time for $R/R_o$ to reach zero, the error is less than $2\%$ and does not affect the final conclusion.

As second verification, we verify the code by simulating the convective flow. \cite{musong2016} had used the IBM to study the heat transfer problem for an isolated isothermal sphere at various Grashof numbers $Gr=g\beta_T\Delta_Td^3/\nu^2$ and Prandtl numbers $Pr=\nu/\kappa$, where $\kappa$, $\beta_T$, $d$ and $\Delta_T$ are the thermal diffusivity, thermal expansion coefficient, diameter of the sphere and the temperature difference between the heated surface and the ambient fluid. We modified our droplet dissolution code to deal with the heat transfer problem and compared with their result at $Gr=100$ and $Pr=0.72$. Figure \ref{fig:2} (b) shows how the normalized heat flux (characterized by Nusselt number $Nu$ defined as the total heat flux across the surface of the sphere over the heat flux in the case of quiescent fluid) changes with time $t$. It can be seen that both the data taken from \cite{musong2016} (black squares) and our numerical results (blue curve) agree with each other. The agreement is not only for the values after reaching the statistical steady state but also for the temporal evolution of $Nu$ over the entire heat transfer process.

\begin{figure}
  \centerline{\includegraphics[width=1.0\textwidth]{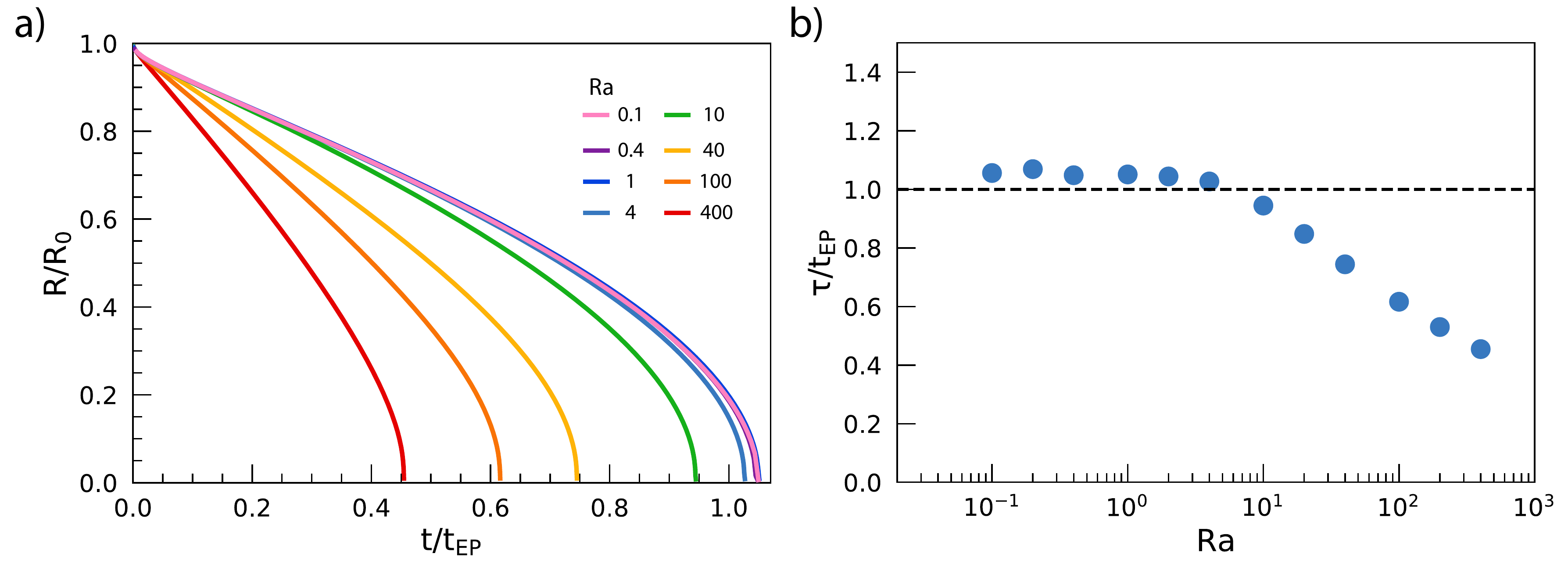}}
  \caption{(a) Time series for the radius of droplet $R(t)$ for different $Ra$ where the $R$ and $t$ are normalized by the initial droplet radius $R_0$ and the droplet lifetime estimated by Epstein-Plesset theory $t_{EP}$. (b) Lifetime of the droplet $\tau$ normalized by $t_{EP}$ versus the Rayleigh number $Ra$. As discussed in Section \ref{sec:verif}, the little deviation from $ t_{EP}$ for small $Ra$ cases is due to the grid resolution issue because the droplet becomes too small at the final stage of dissolution.}
\label{fig:3}
\end{figure}

\begin{figure}
  \centerline{\includegraphics[width=0.6\textwidth]{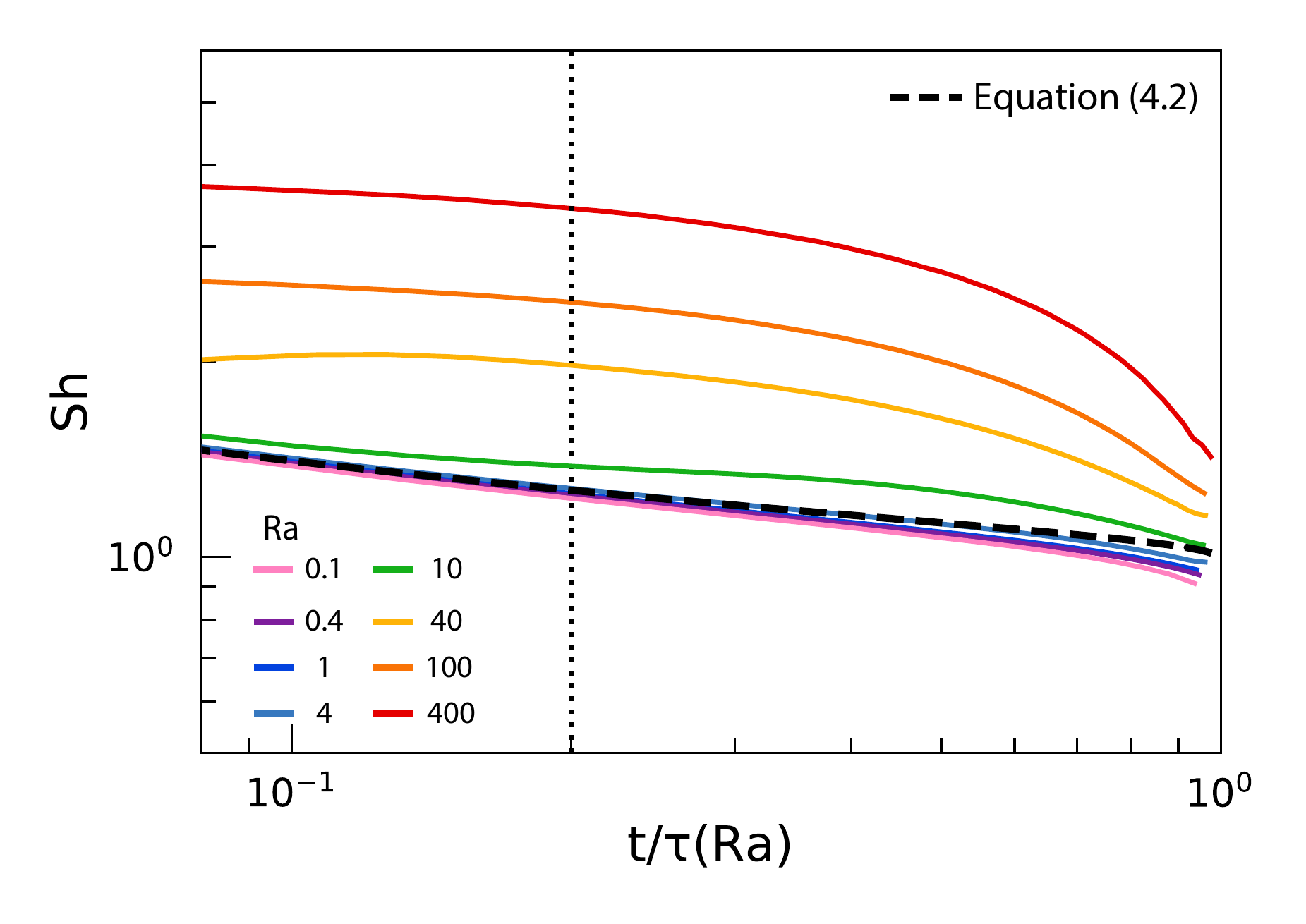}}
  \caption{Instantaneous Sherwood number $Sh(t)$ versus the normalized time $t/\tau(Ra)$ for different $Ra$ where $\tau(Ra)$ is the droplet lifetime for the corresponding $Ra$. The dashed line corresponds to equation (\ref{eq:Sh3}). The vertical dotted line indicates the time instant for the $Sh_{inst}$ shown in figure \ref{fig:5}.}
\label{fig:4}
\end{figure}

\section{Convective effects for single droplet dissolution}\label{sec:resulta}
In this section we first show how the radius of a single surface droplet changes in time for different $Ra$. Figure \ref{fig:3}(a) shows the normalized radius $R(t)/R_0$ versus the normalized time $t/t_{EP}$ for various $Ra$, where $R_0$ and $t_{EP}$ represent the initial droplet radius and the reference lifetime based on the EP theory. For the cases of $0.1\leq Ra\leq 4$, the curves almost collapse onto a single curve, and $R(t)/R_0$ drops to zero when $t \simeq t_{EP}$. It suggests that the droplet dissolution is purely diffusive and we regard those values of $Ra$ as small. However, when $Ra$ increases to $10$, buoyancy force becomes significant as indicated by the curve (green curve) being below the collapsed one at small $Ra$. When $Ra$ further increases from $10$ to $400$, the lifetimes of the droplet are shortened progressively as shown in figure \ref{fig:3}(b) due to the increasing importance of the buoyancy force. For our largest explored $Ra(=400)$, the lifetime of the droplet even becomes half of $t_{EP}$, i.e., half of what it would be for pure diffusion.

\begin{figure}
  \centerline{\includegraphics[width=1.0\textwidth]{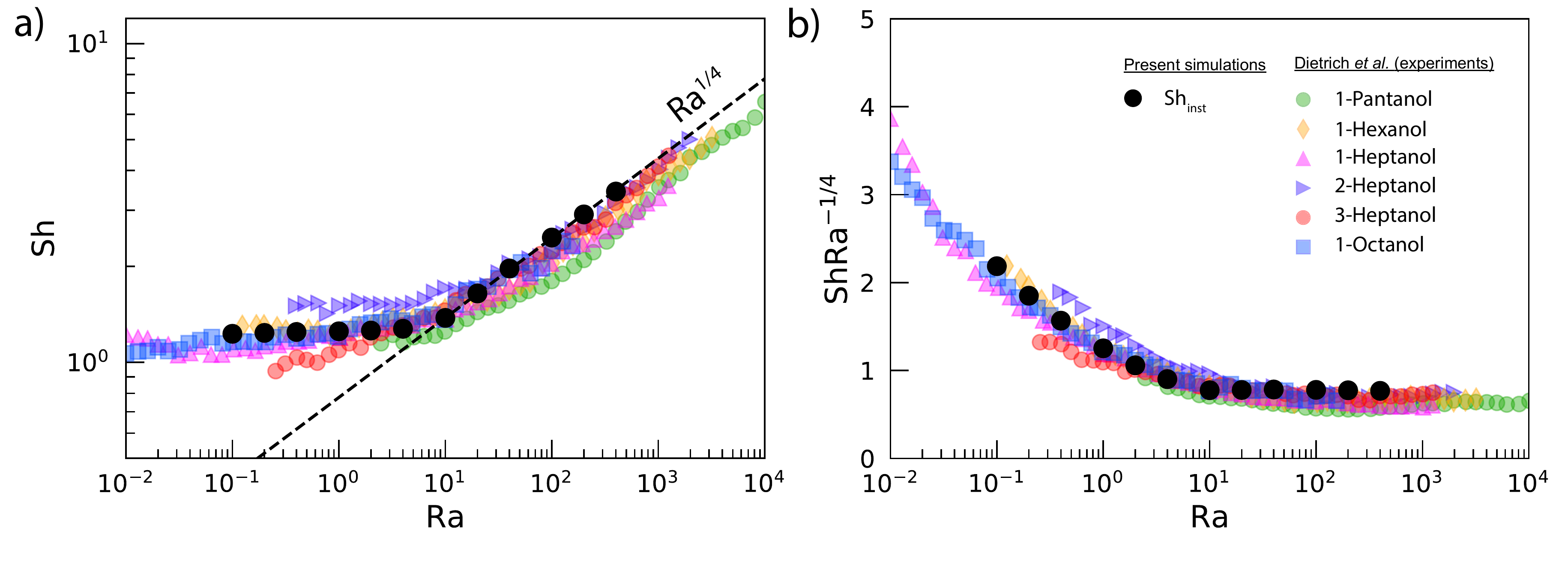}}
  \caption{(a) Sherwood number $Sh$ versus the Rayleigh number $Ra$. For the numerical results, Sherwood number is defined at the instant when the $Sh$ curve is still relatively flat as shown in figure \ref{fig:4}, which is represented by $Sh_{inst}$. For details, we refer to the main text. (b) $Sh$ compensated with $Ra^{1/4}$ versus $Ra$. The experimental data from \cite{dietrich2016role} have also been included.}
\label{fig:5}
\end{figure}

Another important quantity to be examined is the mass flux which in dimensionless form is expressed as the Sherwood number,
\begin{equation} \label{eq:Sh}
Sh=\frac{\langle \dot{m} \rangle_A R}{Dc_s}=\frac{\rho_d}{c_s}\frac{1}{\sqrt{RaSc}}\tilde{R}\frac{d\tilde{R}}{d\tilde{t}}.
\end{equation}
Here $\langle \dot{m} \rangle_A$ is the mass flux averaged over the droplet surface and $Dc_s/R$ is the reference mass flux for the case of the surface droplet (of $90^\circ$ contact angle) dissolving diffusively and quasi-statically. In equation (\ref{eq:Sh}) the expression of $Sh$ is further rewritten to connect it to the (dimensionless) radius shrinkage ${d\tilde{R}}/{d\tilde{t}}$ and the control parameters $Ra$, $Sc$ and $\rho_d/c_s$. 

Given the temporal evolution of the droplet radius as shown in figure \ref{fig:3}(a), the corresponding temporal evolution of $Sh$ can be computed, see figure \ref{fig:4}. Since the lifetimes for different $Ra$ differ a lot from each other, we normalize the time $t$ by the respective lifetime of the droplet $\tau(Ra)$ in each case for better comparison. First, for $0.1\leq Ra\leq 4$, we again observe that the curves collapse with each other. Moreover, $Sh$ changes slightly with time for these cases throughout the entire dissolution process. Using $Ra=0.1$ as an example, the value of $Sh$ is about $1.3$ at $t=0.1\tau$, and then $Sh$ decreases gradually to $1$ until the droplet is completely dissolved. In order to understand this trend, we substitute the analytical solution (\ref{eq:EP}) given by the EP theory into the expression of $Sh$, equation (\ref{eq:Sh}). This gives $Sh(t)$ for the diffusion-dominated case,
\begin{equation} \label{eq:Sh3}
Sh=1+\frac{R}{(\pi D t)^{1/2}}.
\end{equation}
It can be seen that the correction leads to an additive term ${R}/{(\pi D t)^{1/2}}$ to the purely diffusive case under the quasi-static approximation where $Sh=1$. The significance of this term diminishes when $t$ gets larger and $Sh$ approaches $1$ eventually. However, on increasing $Ra$ from $Ra=4$, the expression (\ref{eq:Sh3}) does not hold anymore due to the increasing influence of buoyancy. Upon increasing $Ra$, we observe that the magnitude of $Sh$ becomes larger. Furthermore, take the largest $Ra$($=400$) as an example, one observes that the mass flux remains at a constant value ($Sh \sim 3.8$) over a large portion of the dissolution time, until near the final stage of the dissolution, the value of $Sh$ decreases rapidly. The decrease of the mass flux can be understood as the consequence of the weaker effect of buoyancy due to the smaller effective Rayleigh number caused by the reduced droplet size.

\begin{figure}
  \centerline{\includegraphics[width=1.0\textwidth]{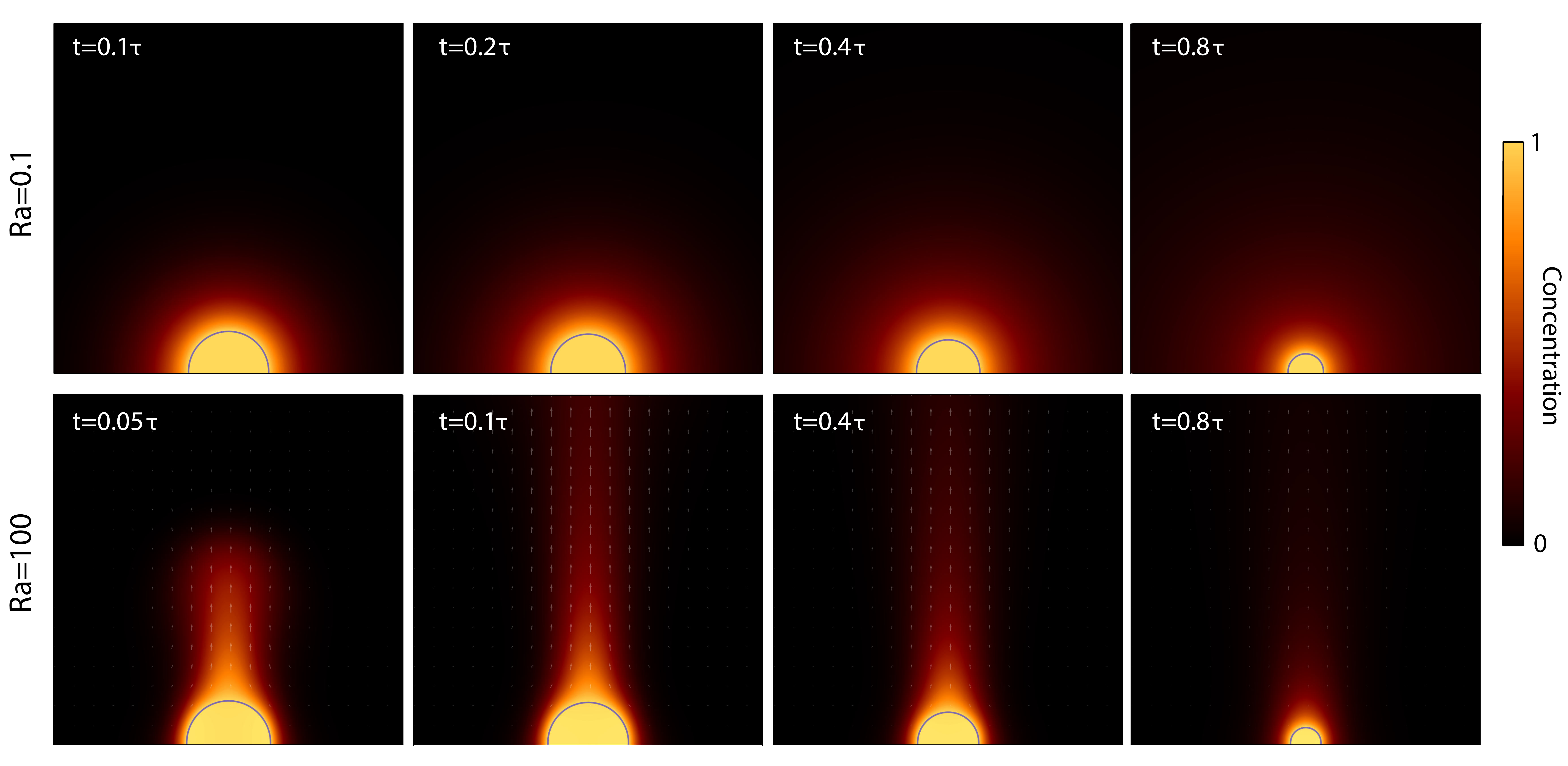}}
  \caption{Instantaneous snapshots for the concentration field together with velocity vectors for $Ra=0.1$ (top) and $Ra=100$ (bottom) in the case of a single surface droplet. The location of this vertical cross-section is taken in the middle of the droplet (also the middle of the domain). The interface of the droplet at different time instants is indicated by the solid line. Movies can be seen in the supplementary material.}
\label{fig:6}
\end{figure}

Next, we examine the dependence of the dimensionless mass flux $Sh$ on $Ra$. Notice from figure \ref{fig:4}, that $Sh$ only slightly decreases for $t\leq0.5\tau$ but then sharply decreases near the final stage of dissolution. It leads us to define an instantaneous Sherwood number $Sh_{inst}$ at the instant when the $Sh$ curve is still relatively flat. Here, the moment of $t=0.2\tau$ is chosen to calculate $Sh_{inst}$ (indicated by the vertical dotted line in figure \ref{fig:4}). Note that our conclusion is insensitive to the choice of the specific time since $Sh$ does not change much near $t=0.2\tau$. In figure \ref{fig:5}(a), it can be seen that on increasing $Ra$, there is a clear transition of $Sh(Ra)$ scaling from a constant to $Ra^{1/4}$. This reflects that the droplet dissolution changes from a diffusively-dominated process to convectively-dominated process because of the increasing significance of buoyancy. To have a better inspection on the scaling change, we further plot $Sh$ compensated with $Ra^{1/4}$ in figure \ref{fig:5}(b) which indeed clearly reveals the $1/4$ scaling exponent.

We now compare our numerical results with the recent experimental results by \cite{dietrich2016role}, who studied the long-chain alcohol droplet dissolving in water and also found an enhanced dissolution rate due to the occurrence of convective flow. In figure \ref{fig:5}, we plot the experimental data points from \cite{dietrich2016role} together with our numerical data for comparison. First, from their experiment, data points from different alcohols have collapsed onto almost the same curve and this curve also displays the change of scaling exponent to $1/4$ on increasing $Ra$. As explained in \cite{dietrich2016role}, this scaling can be understood as follows: For large enough $Ra$ there is a concentration boundary layer developed on top of the droplet surface. The thickness of this boundary layer $\delta_c$ has the Pohlhausen power-law dependence with $Ra$ which is $\delta_c/R \sim Ra^{-1/4}$ \citep{pohlhausen1921naherungsweisen,schlichting2016boundary}. By using $\delta_c$ as the typical length scale for estimating the mass flux, which is $\langle \dot{m} \rangle_A \sim Dc_s/\delta_c$, one can obtain $Sh\sim Ra^{1/4}$. Apart from the scaling change, our numerical results also confirm the value of the transitional Rayleigh number $Ra_t\simeq 12.1$ found in the experiments.

To further characterize the two different dissolution regimes, we compare the respective flow morphologies. Figure \ref{fig:6} shows instantaneous slices of the concentration field taken at the vertical mid-plane. We visualize the time evolution of the concentration by showing the field at different time instants for two different $Ra$. First, for small $Ra$($=0.1$), as shown in the upper row of the figure, the dissolution happens basically through diffusion and one can see that the dissolution rate is almost the same in all directions. On the contrary, this isotropic mass transfer is broken for larger $Ra$, specifically for $Ra \geq 10$. For example at $Ra=100$, as shown in the lower row, the vertical velocity above the droplet strengthens significantly so that the concentration field is mainly displaced upwards rather than sidewards. Near the initial stage of dissolution at $t=0.05\tau$, one can observe the emission of concentration plume from the top of the droplet. When the solute dissolves into water, it results in less dense liquid in the denser surrounding, and such an unstable stratified region leads to the emission of plumes. This mechanism of concentration plume emission is similar to the thermal plume emission in Rayleigh-B\'enard convection which is a classical model for thermal convection with a fluid layer heated from below and cooled from above \citep{shang2003measured,ahlers2009heat}. As the droplet continues to dissolve, a long tail of the plume remains connected to the top of the droplet until the droplet is completely dissolved.

\section{Convective effects for multiple droplet dissolution}\label{sec:resultb}
\begin{figure}
  \centerline{\includegraphics[width=1.0\textwidth]{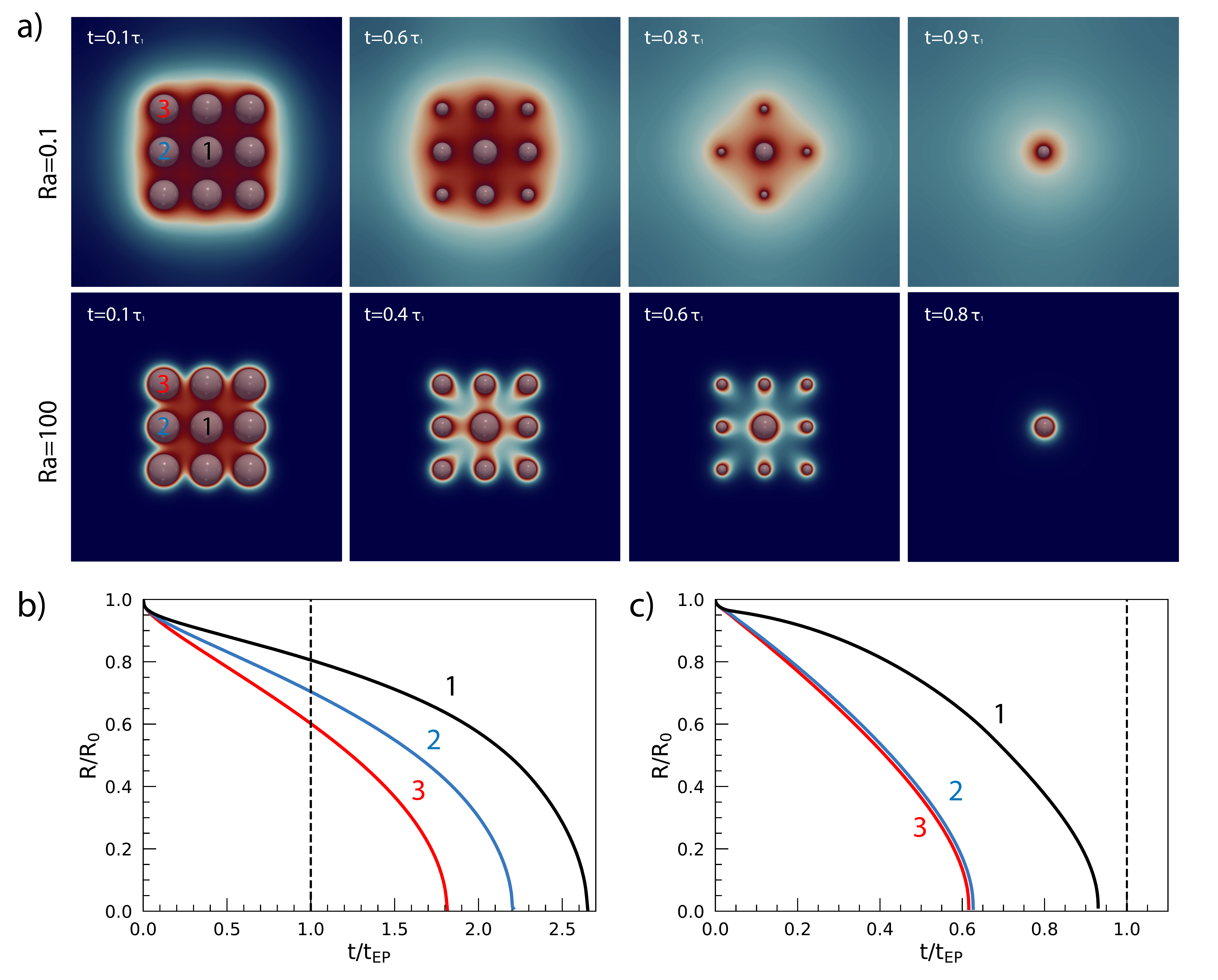}}
  \caption{(a) Top view for the instantaneous concentration fields taking at the layer close to the bottom plate for $Ra=0.1$ and $Ra=100$. To guide the eye, the interfaces of the droplets are also outlined by the grey surfaces. Time series for the normalized radius $R(t)/R_0$ versus the normalized time $t/t_{EP}$ for $Ra=0.1$ in (b) and $Ra=100$ in (c). Here $R_0$ and $t_{EP}$ denote the initial droplet radius and the single droplet lifetime estimated by the EP theory. To denote the droplets at different topological locations, they are indexed with the number $1$, $2$ and $3$ as indicated in (a) for $t=0.1\tau_1$.}
\label{fig:7}
\end{figure}

Given the good agreement with the experimental results, we now extend our numerical study to the case of multiple droplets. Two different multiple droplet configurations are studied, namely $2\times2$ and $3\times3$ droplet arrays. To compare the different dissolution dynamics in the diffusion-dominated and convection-dominated regimes, figure \ref{fig:7}(a) shows the top view (cutting near the bottom plate on which the droplets are placed) of the concentration fields for $Ra=0.1$ and $Ra=100$ using the $3\times3$ array as an example. To also have a quantitative comparison, figures \ref{fig:7}(b,c) show the normalized radius $R/R_0$ versus the normalized time $t/t_{EP}$, where $t_{EP}$ is the lifetime estimated by the EP theory for a single droplet.

For $Ra=0.1$ the collective dissolution leads to much faster accumulation of solute among the droplets as compared to the single droplet dissolution. This is the so-called shielding effect \citep{laghezza2016collective,carrier2016evaporation,bao2018flow,michelin2018collective,wray2019competitive}. The existence of the neighbouring droplets tends to lower the concentration gradient experienced by all the droplets and results in a decreased dissolution rate. Another feature of the shielding effect is that the dissolution of the multiple droplets follows the sequence $\tau_3< \tau_2 < \tau_1$ where $\tau_i$ is the lifetime for the i-th droplet (see figure \ref{fig:7}a for $t=0.1\tau_1$ for the locations). Indeed, both the qualitative visualization in figure \ref{fig:7}(a) and the radius time series in figure \ref{fig:7}(b,c) confirm such a sequence of dissolution. Moreover, we show that for all the droplets, they dissolve slower than the single droplet case with pure diffusion.

In contrast, the dissolution pattern changes significantly when convection plays a role. The second row of figure \ref{fig:7}(a) shows the concentration field for $Ra=100$. From the footprint of the concentration field at $t=0.4\tau_1$, it shows that the solute tends to flow towards the central droplet. Apart from the change in the concentration distribution, figure \ref{fig:7}(c) further shows that the dissolution time of droplet 3, $\tau_3$, becomes comparable to that of droplet 2, $\tau_2$. This feature is non-trivial and opposes the expectation from the shielding effect.

\begin{figure}
  \centerline{\includegraphics[width=1.0\textwidth]{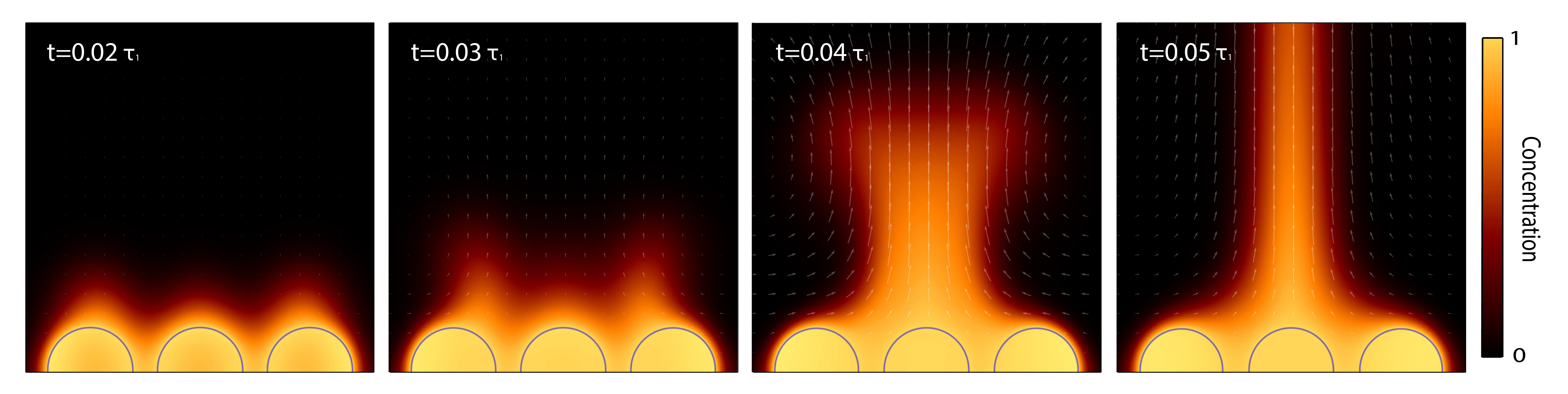}}
  \caption{Instantaneous snapshots for the concentration field together with velocity vectors in the vertical middle plane for $Ra=100$ in the case of multiple droplets of $3\times3$ droplets array. Note that the middle plane is cutting through the center of the droplets 1 and 2. Snapshots at different time instants indicates the formation of single, larger plume from individual plumes. The time $\tau_1$ represents the lifetime of the central droplet (droplet 1). A movie of this process can be seen in the supplementary material.}
\label{fig:8}
\end{figure}

To understand this counter-intuitive result, we thus explore the morphological changes in the flow caused by the significant influence of convection for multiple droplets. Figure \ref{fig:8} visualizes the concentration field at the vertical mid-plane for $Ra=100$. At the initial stage of dissolution ($t=0.02\tau_1$), we observe the plumes emitted mainly from the two side droplets. For the central droplet, the concentration gradient is largely diminished due to the existence of the neighbouring droplets. Therefore, at $t=0.03\tau_1$, we find that the upward velocity above the central droplet is weaker than that above the side droplets. However, instead of just moving upward, the concentration plumes tend to merge together above the central droplet. Eventually, the merging event results in a single larger plume moving vertically upward at $t=0.04\tau_1$. Finally, at $t=0.05\tau_1$ the narrow tail of the plume is maintained and this morphology remains for the rest of the dissolution process until the droplets are completely dissolved. 

So far we have revealed that the plumes need not be individual but that they can interact with each other, leading to a new mechanism for collective droplet dissolution through merging of plumes. It somewhat mimics the daily life example (at the days of Michael Faraday) of two merging flames from two nearby candles: As the fluid in the middle of the two candles receives the strongest heating from the two flames, there is stronger updraft between the two candles and the merged flame can reach higher position. Likewise, a more energetic merged plume can form for multiple droplets dissolution which enhances the mass transfer. We cite Michael Faraday's \emph{Chemical History of a Candle} \citep{faraday1861course}: ``\emph{There is no better, there is no more open door by which you can enter into the study of natural philosophy than by considering the physical phenomena of a candle}." Here, we have recognized the similarity between the candle melting and the droplet dissolving in their collective behaviours, and therefore it enlightens the research on droplet dissolution. In the analogous case---a bubble---\cite{lhuissier2012bursting} have also found that rich fluid mechanics can be learnt through studying the bursting bubble similar to the study of the melting candle.

\begin{figure}
  \centerline{\includegraphics[width=1.0\textwidth]{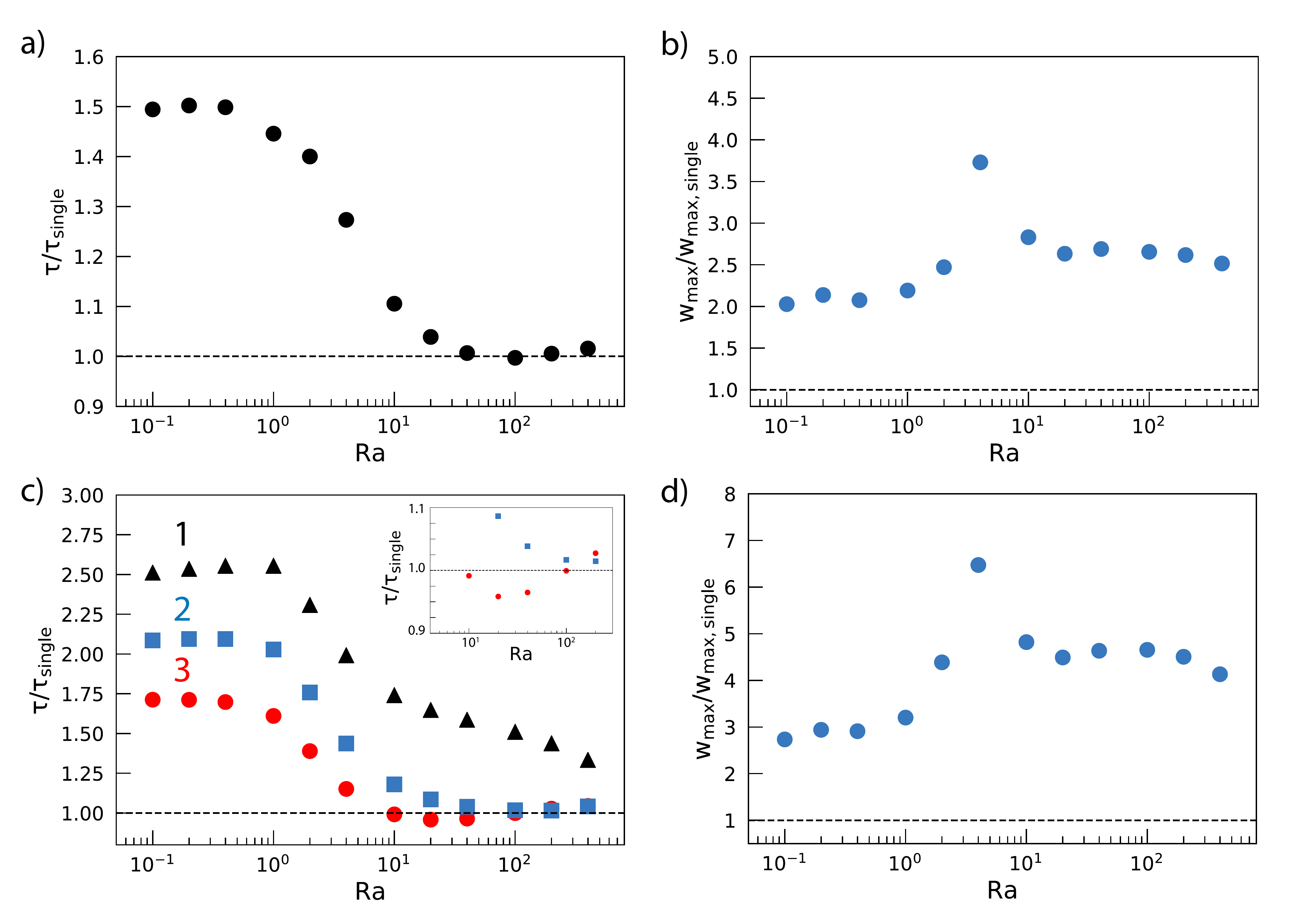}}
  \caption{Normalized droplet lifetime $\tau/\tau_{single}$ in the case of multiple droplets versus $Ra$ for $2\times2$ droplet array in (a) and $3\times3$ droplet array in (c). $\tau_{single}$ represents the lifetime in the case of single droplet. The indices represents the droplets from different locations as indicated in figure \ref{fig:7}(a). The inset in (c) shows that the multiple droplet lifetime can be \emph{shorter} than the single droplet lifetime for large enough $Ra$. It also shows the minimal normalized lifetime at $Ra=20$. Maximum vertical velocity $w_{max}$ normalized by that in single droplet case $w_{max,single}$ versus $Ra$ for the $2\times2$ droplet array in (b) and the $3\times3$ droplet array in (d). Both show a pronounced maximum around $Ra=4$.}
\label{fig:9}
\end{figure}

To demonstrate the effect of the merging event on the lifetime of the droplets, figure \ref{fig:9} (a,c) shows the normalized lifetime $\tau/\tau_{single}$ versus $Ra$ for $2\times2$ and $3\times3$ droplet arrays. Here the lifetimes $\tau$ have been normalized by the respective reference value $\tau_{single}(Ra)$ which is the $Ra$-dependent lifetime corresponding to the single droplet case. Besides, we also plot the maximum vertical velocity $w_{max}$ at the mid-height versus $Ra$ for both arrays in figure \ref{fig:9} (b,d). Again, the vertical velocity has been normalized by the value obtained from the respective single droplet case $w_{max,single}(Ra)$, which also depends on $Ra$ of course.

\begin{figure}
  \centerline{\includegraphics[width=0.6\textwidth]{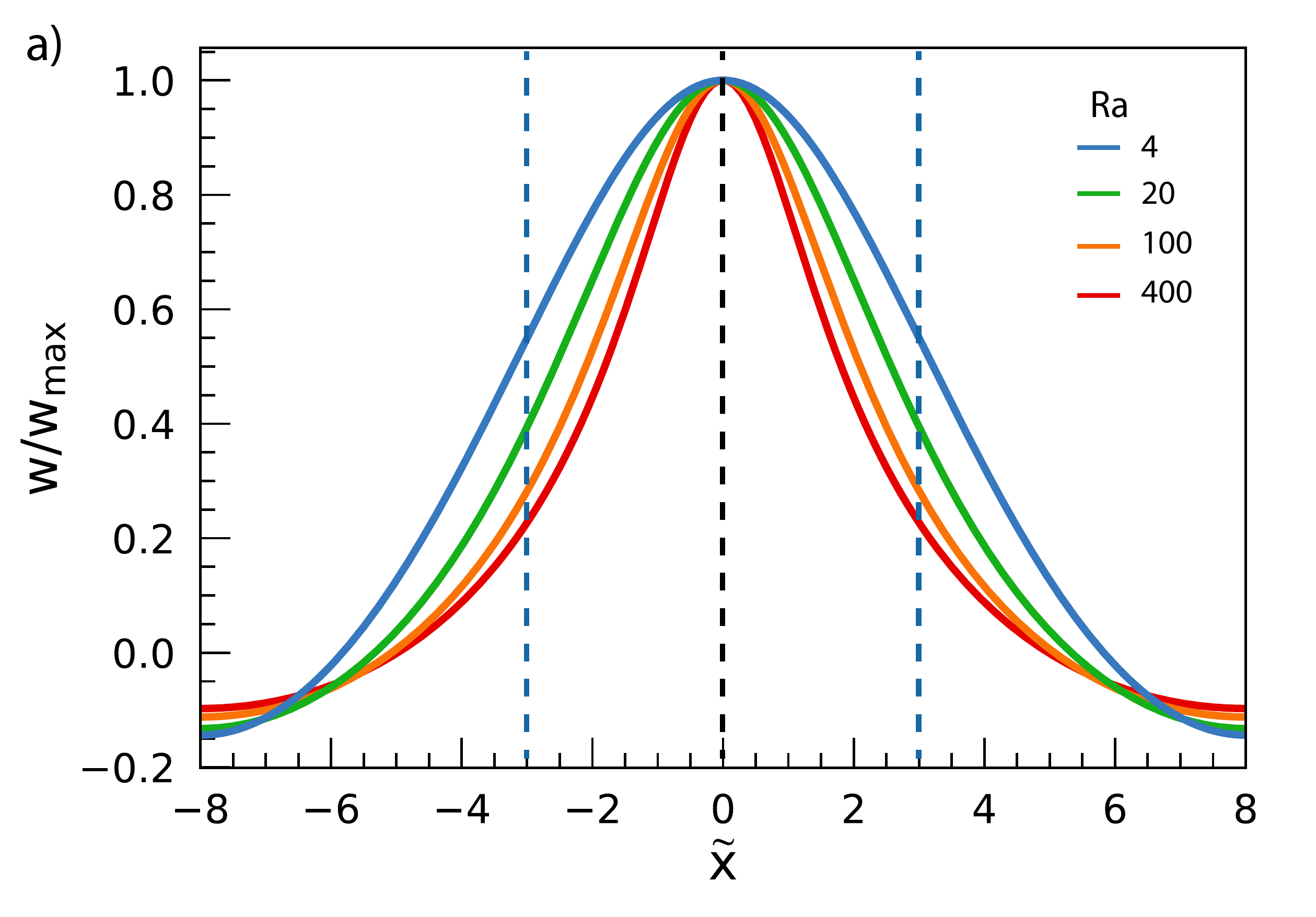}}
  \caption{Horizontal profiles of the normalized vertical velocity $w/w_{max}$ for various $Ra$ where $w_{max}$ is the maximum value of the respective profile. The dimensionless horizontal coordinate is represented by $\tilde{x}$ and we plot the vertical dashed line to show the locations of droplet 2 (blue) and droplet 1 (black).}
\label{fig:10}
\end{figure}

For the $2\times2$ droplet array, all the four droplets are topologically equivalent and therefore figure \ref{fig:9} (a) only shows one set of data. When $Ra \leq 1$ the normalized dissolution time $\tau/\tau_{single}$ is insensitive to the change of $Ra$ and the multiple droplets dissolve slower than the single droplet with $\tau=1.5\tau_{single}$. However, $\tau/\tau_{single}$ decreases with increasing $Ra$ when $Ra$ becomes larger than $1$. As $Ra$ increased up to around $40$, the lifetimes of the multiple droplets become comparable to that of the single droplet. With further increasing $Ra$, the value of $\tau/\tau_{single}$ again becomes insensitive to the change of $Ra$ and stays at around one. This reduction in the lifetime $\tau$ compared to $\tau_{single}$ can be explained by the enhanced vertical velocity shown in figure \ref{fig:9} (b). Due to the merging of the concentration plumes, the maximum velocity $w_{max}$ is considerably larger than that of the single droplet case $w_{max,single}$.

Likewise, for the $3\times3$ droplet array, the trend of $\tau/\tau_{single}$ is similar to that of $2\times2$ array but there is a stronger collective effect. In this $3\times3$ array, there are three topologically different droplets. We display their lifetime versus $Ra$ in figure \ref{fig:9}(c). From that, we can basically classify three different regimes based on the slopes of the curves: In regime I where $Ra\leq1$, the normalized lifetime remains unchanged with increasing $Ra$. In this parameter range, the droplet dissolution is still limited by diffusion and the shielding effect dominates the dissolution process. In the range $1\leq Ra\leq10$ (regime II), we recall that the single droplet dissolution within this $Ra$ range should be diffusion-dominated. However, here we observe the decrease of the normalized lifetime with increasing $Ra$, which reflects the increased influence of the buoyancy force due to the collective droplets. Indeed, figure \ref{fig:9}(d) also shows the significant enhancement in the vertical velocity in this $Ra$ range. In regime III ($Ra\geq10$), we observe that the normalized lifetime of the outermost droplet 3 has reached a plateau where $\tau/\tau_{single}$ stays at around one. To inspect the behaviour of $\tau/\tau_{single}$ at the transition from regime II to III in more detail, we zoom into the region around $Ra=10$ as shown in the inset of figure \ref{fig:9}(c). We observe that there is an optimal case at $Ra=20$ where the lifetime of droplet 3 becomes even shorter (about 5\%) than that in the case of a single droplet. This in fact holds for the whole range $10\leq Ra\leq100$.

\begin{figure}
  \centerline{\includegraphics[width=1.0\textwidth]{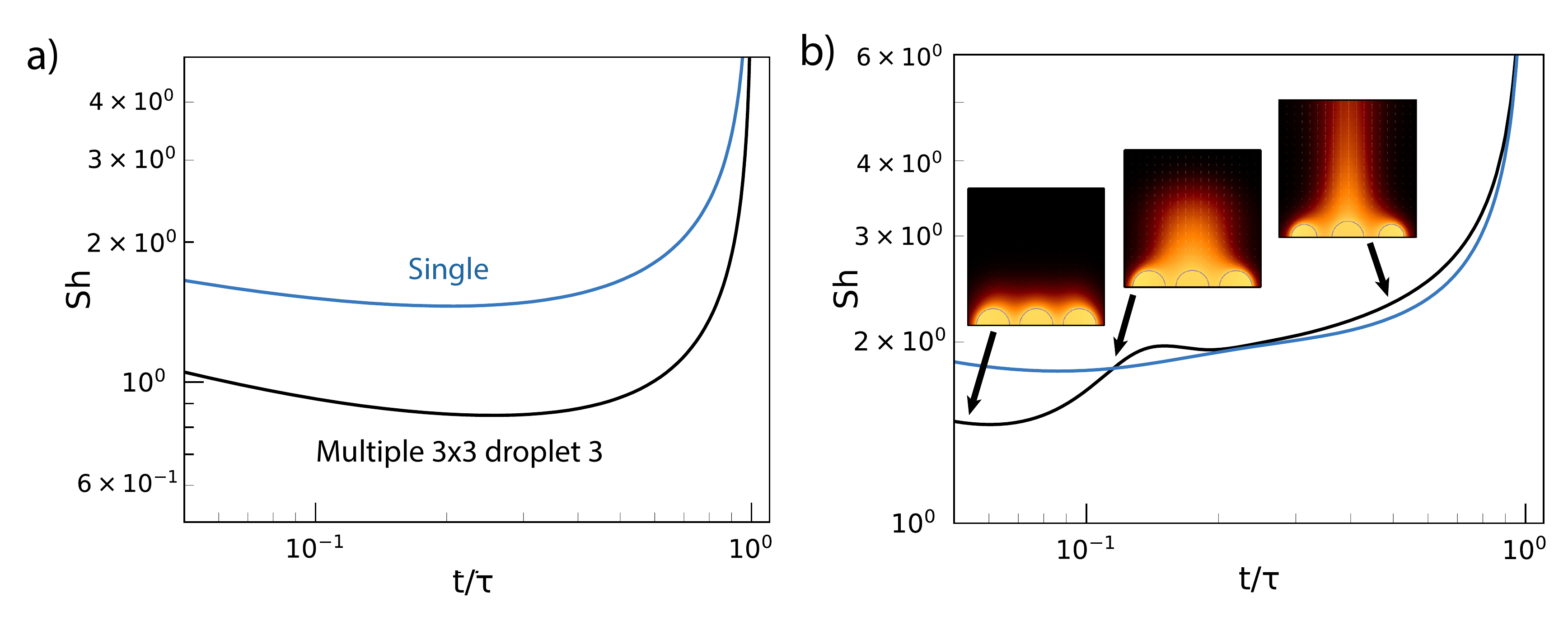}}
  \caption{Time series of Sherwood number $Sh$ versus the normalized time $t/\tau$ where $\tau$ is the lifetime of the droplet at $Ra=0.1$ in (a) and $Ra=20$ in (b). For each $Ra$, the blue curve represents the case of single droplet while the black curve represents the outermost droplet (droplet 3) in the case of $3\times3$ droplet array. In (b),  the vertical cross-section of the concentration fields at different time instants is also shown. Depending on the moment in time, the collective dissolution is either stronger or weaker than that of the isolated droplet.}
\label{fig:11}
\end{figure}

Given that larger $Ra$ represent a relatively stronger buoyancy effect, the above observation indeed raises the question on why there is an optimal $Ra$ at which there is a maximum reduction in the droplet lifetime compared to the single droplet case. We explain it by showing the horizontal profiles of the vertical velocity $w$ (normalized by the maximum vertical velocity $w_{max}$) taken at the mid-height in figure \ref{fig:10}. The profiles are taken at the instant when droplet 2 is half of its initial radius. For all $Ra$, the profiles exhibit a maximum at the center ($\tilde{x}=0$ where the entire horizontal extend ranges from $\tilde{x}=-8$ to $\tilde{x}=8$) and the profiles are symmetric about the central line. A key feature is that when $Ra$ increases from $4$ to $400$, the profiles become narrower as noticed by the half maximum of the profiles. The consequence is that at the location of the droplet 2, which is either $\tilde{x}=-3$ or $\tilde{x}=3$, the value of $w/w_{max}$ actually decreases with increasing $Ra$. This suggests that although the effect of buoyancy is stronger at larger $Ra$, the vertical velocity experienced by the edge droplets can be diminished due to the shrinkage in width of the upward-moving merged plume. 

To better understand the optimal case, for which the normalized dissolution time is minimal as function of $Ra$, we examine the time series of $Sh$. For comparison, we begin with the time series for the case of $Ra=0.1$ in figure \ref{fig:11}(a). It shows that for the outermost droplet (droplet 3), the value of $Sh$ during the entire dissolution process is lower than that of the single droplet dissolution, thanks to the shielding effect. However at the optimal case of $Ra=20$, figure \ref{fig:11}(b) shows that the value of $Sh$ for the outermost droplet is not always smaller than that of the single droplet case: First, when $t$ is below $0.1 \tau$, the value of $Sh$ for droplet 3 is lower than that of the single droplet. By the corresponding concentration field over that period of time, one can see that the individual concentration plumes just emit from the droplets without merging at this early stage. However, there is a crossover around $t=0.1\tau$ where the individual plumes are observed to just merge into a single plume. After that, $Sh$ for the outermost droplet remains larger than that for the single droplet case. The result again confirms that it is the merging of plumes leading to the enhancement of the dissolution rate. 

Note that \cite{laghezza2016collective} have also experimentally observed the enhancement of the mass flux for collective and convective dissolution. Thanks to our numerical work, this enhancement can now be linked to the merging of the plumes.

\section{Concluding remarks and outlook} \label{sec:conc}
In summary, we numerically modelled and investigated convective droplet dissolution over a wide range of $Ra$ from $0.1$ to $400$ with $Sc$ being fixed at $1200$ and $c_s/\rho_d$ fixed at $0.027$ (representing 1-pentanol in water). For all our explored cases, we consider the constant contact angle dissolution mode with contact angle being fixed at $90^\circ$. As the starting point, we verified our code for the pure diffusive droplet dissolution by comparing with the analytical results by \cite{epstein1950stability} and \cite{popov2005evaporative}. We then provided further verification to show our proper implementation of the convective term in our code by comparing to the heat flux data in a heat transfer problem. Then we used this numerical code to simulate droplet dissolution for both the single droplet and multiple droplets scenarios. 

For a single droplet, we showed that the Sherwood number $Sh$ stays at around one regardless of $Ra$, provided that $Ra$ is smaller than $10$. However, $Sh$ undergoes a transition to $Sh \sim Ra^{1/4}$ when $Ra$ is above $10$. Our numerical results agree with the previous experimental results by \cite{dietrich2016role} for single droplet dissolution, in which the transition from a constant value to $Sh \sim Ra^{1/4}$ was also found at the same $Ra_t \simeq 10$. Moreover, we gained insight into the change in the flow morphologies by comparing the concentration fields in the different regimes. An essential feature of the convective regime $Sh\sim Ra^{1/4}$ is that there is a clear emission of concentration plumes above the droplet which carries large amount of solute away from the droplet. Our results thus illustrated, from both the $Sh$ behaviour and the flow morphologies, how, with increasing $Ra$, the dynamics of droplet dissolution changes from diffusion-dominated to convection-dominated.

When we extended the geometry to multiple droplets, richer phenomena could be observed. With multiple droplets, the traditional view was that the shielding effect can lead to the large suppression of mass flux due to the smoothened concentration gradient around the droplets. However, the basis of the shielding effect is that the diffusion dominates the dissolution process. Here, with the significant role of convection for large $Ra$, we first showed that the outermost and the second outermost droplet (in $3\times3$ droplets array) have comparable lifetimes which opposes the view of shielding effect. Thanks to the numerical simulations, we further revealed that the concentration plumes can merge into a large, single plume which is the mechanism leading to the collective enhancement of droplet dissolution. With the help of plume merging, the magnitude of vertical velocity is largely increased and the dissolution time for the outermost droplet can be shorter than that of a single droplet by $5\%$ (at $Ra=20$) for our explored parameter range. Based on qualitative experimental observations, \cite{laghezza2016collective} had also reported the enhanced mass flux for multiple droplet dissolution. Here, we understand this enhancement by linking it to the newly found mechanism---plume merging. Another non-trivial result is the existence of an optimal $Ra$. We have provided an explanation by showing that the updraft associated with the large plumes becomes narrower for larger $Ra$. It eventually limits the mass flux of the droplets near the edge as those droplets are less affected by the upward-moving merged plume.

To the best of our knowledge, our study is the first of its kind to provide a detailed physical quantification of the convective collective droplet dissolution problem using numerical simulations. The present study reveals a variety of physical effects thanks to the interplay between the two mechanisms, namely the shielding effect and the merging of concentration plumes. Our findings have thus provided a more comprehensive picture of the collective behaviour of multiple droplets dissolution.

Many questions remain open. E.g., how does the separation between multiple droplets influence the relative strength of the two mechanisms? How do things change for different contact angle $\theta\neq 90^\circ$, or even for different dissolution mode, namely for the constant contact radius mode rather than the constant contact angle mode as employed here? As we have demonstrated some non-trivial and at first sight counter-intuitive results in collective and convective droplet dissolution, it is clearly worthwhile to further explore the parameter space of this system.

\section*{Acknowledgements}
We greatly appreciate the valuable dissuasions with Xuehua Zhang, Andrea Prosperetti and Emmanuel Villermaux. We acknowledge the support from ERC-Advanced Grant under the project number $740479$. K. L. C. also acknowledges Croucher Foundation for the Croucher Fellowships for Postdoctoral Research. We acknowledge that the results of this research have been achieved using the DECI resource Kay based in Ireland at Irish HPC center with support from the PRACE. We also acknowledge PRACE for awarding us access to MareNostrum at the Barcelona Supercomputing Centre (BSC) under PRACE project number 2017174146 and JUWELS at the J\"ulich Supercomputing Centre. This work was also partly carried out on the national e-infrastucture of SURFsara with the support of SURF Cooperative

\section*{Declaration of Interests}
The authors report no conflict of interest.

\end{document}